\documentclass[pre,superscriptaddress,floatfix,twocolumn]{revtex4}

\usepackage{graphicx}
\usepackage{amssymb}

\newcommand{\pd}[2]{\frac{\partial #1}{\partial #2}}

\newcommand{\vv}[1]{{\bf #1}}

\newcommand{\W}{\Omega}
\newcommand{\w}{\omega}
\renewcommand{\k}{\vv{k}}
\newcommand{\p}{\vv{p}}
\newcommand{\q}{\vv{q}}
\renewcommand{\r}{\vv{r}}

\begin{document}

\title{Nonlocal wave turbulence in the Charney-Hasegawa-Mima equation: a short review}
\author{Colm Connaughton}
\email{connaughtonc@gmail.com}
\affiliation {Centre for Complexity Science, University of Warwick, Gibbet Hill
Road, Coventry CV4 7AL, UK}
\affiliation{Mathematics Institute, University of Warwick, Gibbet Hill
Road, Coventry CV4 7AL, UK}
\author{Sergey Nazarenko}
\email{S.V.Nazarenko@.warwick.ac.uk}
\affiliation{Mathematics Institute, University of Warwick, Gibbet Hill
Road, Coventry CV4 7AL, UK}
\author{Brenda Quinn}
\email{B.E.Quinn@warwick.ac.uk}
\affiliation{Mathematics Institute, University of Warwick, Gibbet Hill
Road, Coventry CV4 7AL, UK}

\date{\today}

\begin{abstract}
Rossby wave turbulence, as modelled by the Charney-Hasegawa-Mima (CHM) equation,
is nonlocal in scale. As a result, the formal stationary Kolmogorov-Zakharov 
solutions of the Rossby wave kinetic equation, which describe local cascades, are 
not valid. Rather the solution of the kinetic equation is dominated by 
interactions between the large and small scales. This suggests an alternative
analytic approach based on an expansion of the collision integral in a small
parameter obtained from scale separation. This expansion approximates the
integral collision operator in the kinetic equation by anisotropic diffusion 
operators in wavenumber space as first shown in a series of papers by Balk, 
Nazarenko and 
Zakharov in the early 1990's. In this note we summarise the foundations of this 
theory and provide the technical details which were absent from the original papers.
\end{abstract}


\maketitle

\section{Introduction to weak wave turbulence in the Charney-Hasegawa-Mima equation}

\begin{figure}
\includegraphics[width=6.5cm] {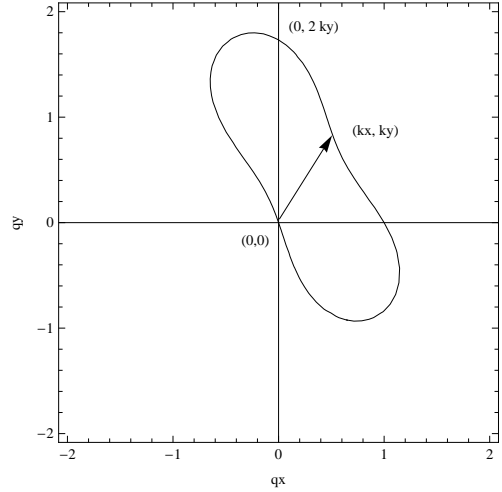}
\caption{\label{fig-resonantManifolds}Typical shape of the resonant manifolds 
defined by Eq.~(\ref{eq-resonantManifolds}).}
\end{figure}

In the limit of weak nonlinearity, the statistical evolution of CHM turbulence can
be described by the theory of weak wave turbulence. See \cite{ZLF92,NNB01} for
a review of the theory. The principal result of this theory is the fact that that
the wave spectrum, $n_\k$, evolves according to the wave kinetic equation:
\begin{eqnarray}\nonumber
        \frac{\partial n_\vv{k}}{\partial t}
&=&
4 \pi \!\! \int   \! \left| V^\vv{k}_{\q\,\r}
\right|^2 \delta(\k - \q - \r) \delta(\omega_\k -\omega_\q - \omega_\r) \! \times\\
\nonumber
&       & \!\!\!\!\!\!\!\!\!\!\! \!\!\!\!\!
\left[ n_\q n_\r - n_\k n_\q \hspace{.5mm} \mathrm{sgn}  (\omega_\k\omega_\r)
- n_\k n_\r \hspace{.5mm} \mathrm{sgn} (\omega_\k\omega_\q) \right]
 d \q d \r,
\\ \label{eq-KE}
&       & 
\end{eqnarray}
where 
\begin{equation}
\label{eq-omega}
\w_\k = -\frac{\beta\,\rho^2\,k_1}{1+\rho^2\,k^2}
\end{equation}
and
\begin{equation}
\label{eq-V}
V^\vv{k}_{\q\,\r}
=  \frac{i}{2} \rho^2\,\sqrt{\beta\left| k_r q_1 r_1\right|} \left( \frac{q_2}{1+\rho^2 q^2} + \frac{r_2}{1+\rho^2 r^2}-\frac{\rho^2k_2}{1+\rho^2k^2} \right).
\end{equation}
Throughout these notes we shall use boldface text to denote vectors in ${\mathbb R}^2$, regular text to denote their magnitudes and subscripts to denote their
components. Thus, for example,
 $\k = (k_1, k_2)$ and $k = \left|\k\right| = \sqrt{k_1^2+k_2^2}$.
The nonlinear interaction coefficient, $V^\k_{\q\,\r}$ has several symmetries
which will be useful later:
\begin{eqnarray}
\nonumber V^{(k_1,k_2)}_{(q_1,q_2)\,(r_1,r_2)} &=& V^{(k_1,k_2)}_{(r_1,r_2)\,(q_1,q_2)} = -V^{(r_1,r_2)}_{(k_1,k_2)\,(-q_1,-q_2)}\\
\nonumber&=& -V^{(k_1,-k_2)}_{(q_1,-q_2)\,(r_1,-r_2)} = V^{(-k_1,k_2)}_{(q_1,q_2)\,(r_1,r_2)}\\
&=& V^{(k_1,k_2)}_{(-q_1,q_2)\,(r_1,r_2)} = V^{(k_1,k_2)}_{(q_1,q_2)\,(-r_1,r_2)}.
\label{eq-Vsymmetries}
\end{eqnarray}
According to Eq.~(\ref{eq-KE}), exchange of energy is only possible among
triads of modes, $(\k,\q,\r)$, which satisfy the resonance conditions:
\begin{eqnarray}
\label{eq-resonantManifolds}\k&=&\q + \r\\
\nonumber \w_\k &=& \w_{\q} + \w_{\r}.
\end{eqnarray}
For each given $\k$, the modes $\q$ permitted to interact with $\k$ lie on
a one-dimensional curve in the $(k_1, k_2)$ plane known as the 
resonant manifold of the mode $\k$ (the third member of the triad is given by
$\r = \k-\q$). The resonant manifolds typically have the shape shown in
Fig.~\ref{fig-resonantManifolds}.

The kinetic equation, Eq.~(\ref{eq-KE}), becomes scale invariant only in the
limits, $k \gg \rho^{-1}$ and $k\ll \rho^{-1}$. In these limits, it is possible to 
formally find stationary Kolmogorov-Zakharov (KZ) solutions of Eq.~(\ref{eq-KE}) 
describing turbulent cascades of energy and enstrophy. The inverse cascade spectrum,
describing the transfer of energy from small scales to large, turns out to
be non-local meaning that the RHS of Eq.~(\ref{eq-KE}) diverges on this spectrum
so that it is not a consistent stationary solution. It is worth stressing
that the fact that the ``usual'' KZ spectrum is not realisable does not
necessarily invalidate the use of the kinetic equation itself.  Rather, the 
non-locality of the KZ solution leads one to expect that the transfer of energy 
from small scales to large in the CHM equation proceeds by a direct interaction 
between the large and small scales rather than via a scale-local cascade process. 
This is what is meant by non-local turbulence. In this section, we provide a 
detailed description of this process. This analysis was originally 
presented in \cite{BNZ1990,BNZ1990B} and is reproduced here in slightly
expanded form for the sake of clarity and completeness.

The x-component of the phase speed of a wave having wavenumber $\k$ is
\begin{equation}
\label{eq-phaseSpeed}
c_\k = \frac{\w_\k}{k_1} = -\frac{\beta\,\rho^2}{1+\rho^2\,k^2}.
\end{equation}
Notice that all waves have a phase velocity to the left. Furthermore for large
scale waves, $k \lg \rho^{-1}$, $c_\k \to - \beta \rho^2 = -v^*$ and 
all waves travel at the same speed, known as the drift velocity:
\begin{equation}
\label{eq-driftVelocity}
v^* = \beta \rho^2.
\end{equation}
We are interested in describing a scenario in which small scale turbulence 
evolves by direct interaction with large scale waves. Since the large scale
waves are all travelling at the drift velocity in the x-direction, it is natural 
to work in a frame moving in the x-direction with the drift velocity, 
$-\beta\,\rho^2$. In this frame, using Eq.~(\ref{eq-phaseSpeed}), the phase
speed of mode $\k$ changes to:
\begin{displaymath}
c_\k \to c_\k + \beta\,\rho^2 = \frac{\beta\,\rho^4\,k^2}{1+\rho^2k^2}.
\end{displaymath}
The frequencies are Doppler-shifted accordingly:
\begin{equation}
\label{eq-DopplerShift}
\w_\k = c_\k\,k_1 \to \beta\,\rho^2\,k_1 + \w_\k.
\end{equation}
We shall use $\W_k$ to denote the frequency in the moving frame:
\begin{equation}
\label{eq-W}
\W_\k = \rho^2\,k_1 + \w_\k = \frac{\beta\,\rho^4\,k^2\,k_1}{1+\rho^2k^2}.
\end{equation}
In the moving frame the kinetic equation, Eq.~(\ref{eq-KE}), becomes
\begin{eqnarray}\nonumber
        \frac{\partial n_\vv{k}}{\partial t}
&=&
4 \pi \!\! \int   \! \left| V^\vv{k}_{\q\,\r}
\right|^2 \delta(\k - \q - \r) \delta(\omega_\k -\omega_\q - \omega_\r) \! \times\\
\nonumber
&       & \!\!\!\!\!\!\!\!\!\!\! \!\!\!\!\!
\left[ n_\q n_\r - n_\k n_\q \hspace{.5mm} \mathrm{sgn}  (\omega_\k\omega_\r)
- n_\k n_\r \hspace{.5mm} \mathrm{sgn} (\omega_\k\omega_\q) \right]
 d \q d \r, \\ 
\label{eq-KE2} &       &
\end{eqnarray}
Note from Eq.~(\ref{eq-DopplerShift}) that the change from $\w_\k\to\W_\k$
does not change the resonant manifolds, Eqs.~(\ref{eq-resonantManifolds}),  due to 
the presence of the momentum delta function, $\delta(\vv{k} - \vv{k}_1 - \vv{k}_2)$.

Let us now consider small scale turbulence with typical wave-vector $\k$. From 
Fig.~\ref{fig-resonantManifolds}, we can see that the resonant manifold containing
the set of wave-vectors, $\k_1$, permitted to interact with $\k$ typically 
intersects the $k_{1}=0$ axis in two places. Setting $k_{1}=0$ in 
Eqs.~(\ref{eq-resonantManifolds}) and doing some calculations shows that 
these points are
\begin{eqnarray}
\label{eq-P1}{\mathrm P}_1 &=& (0,0)\\
\label{eq-P2}{\mathrm P}_2 &=& (0,-2\,k_2).
\end{eqnarray}
If we are interested in non-local interaction between wavenumbers in the
neighbourhood of $\k$ and zonal
flows, then the points ${\mathrm P}_1$ and ${\mathrm P}_2$ will give the dominant
contributions. Interaction with the point ${\mathrm P}_1$ always describes 
interaction with a large scale
zonal flow. The wave number of ${\mathrm P}_2$, on the other hand,  is comparable 
to the $k_2$ component of the small scale turbulence. Interaction with the point ${\mathrm P}_2$ can correspond to interaction with a large or
small scale zonal flow depending on whether the small scale turbulence has
an appreciable  $k_2$ component or not. We focus on the interaction with
large scale zonal flows first.

\section{Nonlocal interaction with large scale zonal flows \label{sec-largeScaleZF}}

Let us introduce a large scale reference scale, $K \ll \rho^{-1}$, and after
integrating out $\r$ from Eq.~(\ref{eq-KE2}), split the right hand side of the
kinetic equation as
\begin{eqnarray}
\label{eq-KE3}\pd{n_\k}{t} &=&  \int_{q<K}\mathrm{Coll}\left[n_\k, n_{\q}, \k,\q\right]\,d\q\\
\nonumber &+&  \int_{q\geq K}\mathrm{Coll}\left[n_\k, n_{\q}, \k,\q\right]\,d\q,
\end{eqnarray}
where the collision integral is:
\begin{eqnarray}
\nonumber
& &\mathrm{Coll}\left[n_\k, n_{\q}, \k,\q\right] = 4 \pi \left| V^\vv{k}_{\q\ \k-\q}
\right|^2 \delta(\W_{\vv{k}} -\W_{\q} - \W_{\k-\q})\\
\nonumber
& & \left[ n_{\q} n_{\k-\q} - n_{\vv{\k}} n_{\q} \mathrm{sgn}  (\w_\k \w_{\k-\q})
- n_{\k} n_{\k-\q}  \mathrm{sgn} (\w_k \w_{\q}) \right]
 d \vv{k}_1\\ 
&&\label{eq-Coll1}
\end{eqnarray}
If we are interested in the dynamics of the small scales, $k\gg K$, then the 
assumption of nonlocality means that we can neglect the second term in 
Eq.~(\ref{eq-KE3}) compared to the first. Further simplifications can be made in
the first term using the fact that $k \gg q$ everywhere in the integrand and
assuming that $n_\k \ll n_{\q}$ everywhere in the integrand. This latter 
inequality means that we are describing a situation where small scale turbulence
is interacting with an intense large scale zonal flow. This discussion
does not tell us how this large scale zonal flow was generated in the first place. 

As $\q\to 0$, $\mathrm{sgn}  (\w_\k \w_{\k-\q})=1$ and we can 
approximate the $n_\k$ dependence in the collison integral as follow:
\begin{eqnarray*}
&&\left[ n_{\q} n_{\k-\q} - n_{\vv{\k}} n_{\q} \mathrm{sgn}  (\w_\k \w_{\k-\q}) - n_{\k} n_{\k-\q}  \mathrm{sgn} (\w_k \w_{\q}) \right]\\
&&\to (n_{\k-\q} - n_\k) \, n_{\q}
\end{eqnarray*}
where we neglect the term $n_{\k} n_{\k-\q}$ describing small-scale
small-scale interactions on the basis that it is $O(n_\k^2)$
as $\q\to 0$ and much smaller than the terms describing small-scale large-scale
interactions. In other words, we assume that $n_\k \ll n_{\q}$.  We therefore
approximate  Eq.~(\ref{eq-KE2}) by
\begin{equation}
\pd{n_\k}{t} =  \int_{q<K} F(\k,\q)\,d\q
\end{equation}
where
\begin{eqnarray}
\nonumber
F(\k,\q) &=& 4\pi \left|V^\q_{\k\!-\!\q\,\k}\right|^2 \,\delta(\W_\k-\W_\q-\W_{\k-\q})\\
\label{eq-F}&&\times n_\q\,\left(n_{\k-\q}-n_\k\right).
\end{eqnarray}
Using the symmetries of $V^\q_{\k\!-\!\q\,\k}$ and $\W_\k$ it can be shown
by direct computation that
\begin{equation}
\label{eq-Fsymm}
F(\k,\q) = -F(\k-\q,\q).
\end{equation}
Thus we can then write:
\begin{eqnarray}
\nonumber \pd{n_\k}{t} &=&   \frac{1}{2}\,\int_{q<K} F(\k,\q)\,d\q - \int_{q<K} F(\k-\q,-\q)\,d\q\\
\nonumber &=& \frac{1}{2}\, \int_{q<K} \left( F(\k,\q) - F(\k+\q,\q) \right) \,d\q \\
\label{eq-KE4} &=& -\frac{1}{2}\,\int_{q<K} d\q\ \q\cdot\nabla_\k F(\k,\q),
\end{eqnarray}
where in the final step we have Taylor expanded $F(\k+\q,\q)$ with respect
to $\q$ in the first argument and neglected terms of $O(q^2)$.

We should now Taylor expand $F(\k,\q)$ with respect to $\q$ in 
Eq.~(\ref{eq-KE4})
To do so, we first Taylor expand the argument of the $\delta$-function. Taylor
expanding $\W_{\k-\q}$ with respect to $\q$ gives:
\begin{displaymath}
\W_{\k-\q} = \W_\k - \q\cdot\nabla_\k\W_\k + O(q^2)
\end{displaymath} 
as $\q\to 0$, where $\nabla_\k = (\partial_{k_1},\partial_{k_2})$ is
the gradient operator in $\k$-space. Thus the difference 
$\W_{\vv{k}} - \W_{\k-\q}$ behaves as $\q\cdot\nabla_\k\W_\k$ as $\q\to 0$.
On the other hand, the remaining term $\W_{\q}$ behaves as $q^2q_1$ as
$\q \to 0$ as can be seen directly from Eq.~(\ref{eq-W}). Thus the latter
can be neglected compared with the former in the limit $\q\to 0$ and
we can replace
\begin{displaymath}
\delta(\W_{\vv{k}} -\W_{\q} - \W_{\k-\q}) \to \delta(\q\cdot\nabla_\k\W_\k)
\end{displaymath}
in Eq.~(\ref{eq-Coll1}). 
Therefore $F(\k,\q)$ in Eq.~(\ref{eq-KE4}) can be approximated as
\begin{eqnarray}
\nonumber F(\k,\q) &\approx& 4\pi \left| V^\vv{k}_{\q\ \k-\q} \right|^2 \delta(\q\cdot\nabla_\k\W_\k) (n_{\k-\q} - n_\k) \, n_{\q}\\
&\approx& -4\pi \left| V^\vv{k}_{\q\ \k-\q} \right|^2 \delta(\q\cdot\nabla_\k\W_\k)\, \q\cdot \nabla_\k\,n_\k,
\label{eq-F2}
\end{eqnarray}
where we have used the fact that 
$n_{\k-\q} - n_\k = -\q\cdot \nabla_\k\,n_\k + O(q^2)$ as $\q\to 0$.
Combining Eq.~(\ref{eq-F2}) with Eq.~(\ref{eq-KE4}), the kinetic
equation for the small scales can be written as an anisotropic diffusion equation 
in 
$\k$-space:
\begin{equation}
\label{eq-diffusion1}
\pd{n_\k}{t} = \pd{}{k_i}\,S_{ij}(q,k_2) \pd{n_\k}{k_j}
\end{equation}
where repeated component subscripts are summed over. The diffusion tensor is
\begin{equation}
\label{eq-diffusionTensor1}
S_{ij}(\k) = 2\pi \int_{q<K} d\q \left| V^\vv{k}_{\q\ \k-\q} \right|^2 \delta(\q\cdot\nabla_\k\W_\k)\, q_i\,q_j\,n_\q.
\end{equation}
Let us now suppose that the large scales are principally supported at scales
$q\ll K$. The reference wavenumber, $K$, can then be extended to infinity in
Eq.~(\ref{eq-diffusionTensor1}) and we can perform the integration with 
respect to $q_1$:
\begin{widetext}
\begin{eqnarray}
\nonumber S_{ij}(\k) &=& 2\pi \int_{-\infty}^\infty d q_1 d q_2  \left| V^\vv{k}_{\q\ \k-\q} \right|^2 \delta(q_1 \pd{\W_\k}{k_1} + q_2 \pd{\W_\k}{k_2})\, q_i\,q_j\,n_\q\\
\nonumber &=& 2\pi \int_{-\infty}^\infty d q_1 d q_2  \left| V^\vv{k}_{\q\ \k-\q} \right|^2 \left|\pd{\W_\k}{k_1} \right|^{-1} \delta(q_1  + \theta_\k q_2 )\, q_i\,q_j\,n_\q\\
\nonumber &=& 2\pi \left|\pd{\W_\k}{k_1}\right| ^{-1} \int_{-\infty}^\infty d q_2  \left[ \left| V^\vv{k}_{\q\ \k-\q} \right|^2  q_i\,q_j\,n_\q \right]_{q_1=-\theta_\k q_2} 
\end{eqnarray}
\end{widetext}
where we have introduced the shorthand parameter
\begin{equation}
\label{eq-theta}
\theta_\k = \frac{\pd{\W_\k}{k_2}}{\pd{\W_\k}{k_1}}
\end{equation}
Some elementary calculations bring the diffusion tensor to the form:
\begin{equation}
\label{eq-diffusionTensor2}
S(\k) = \tilde{S}(\k)\, \pd{\W_\k}{k_1}\, \left(\begin{array}{cc}\theta_\k^2&-\theta_\k\\-\theta_\k&1\end{array} \right).
\end{equation}
where we have introduced the scalar function
\begin{equation}
\tilde{S}(\k) = 2\pi \left(\pd{\W_\k}{k_1}\right)^{-2} \int_{-\infty}^\infty d q_2  \left[ \left| V^\vv{k}_{\q\ \k-\q} \right|^2 \,n_\q \right]_{q_1=-\theta q_2}.
\end{equation}
The matrix in Eq.~(\ref{eq-diffusionTensor2}) is singular. It has eigenvalues
$0$ and $1+\theta_\k^2$ with corresponding eigenvectors $(\theta_\k^{-1}, 1)$ and 
$(-\theta, 1)$. This suggests that at any point in  $\k$-space, diffusion only
occurs in the direction $(-\theta_\k, 1)$ corresponding to the positive 
eigenvalue. Eq.~(\ref{eq-diffusion1}) should then describe diffusion along
one-dimensional curves in $\k$-space. In order to show this and to find these
curves, we need to find a change of variables which reduces 
Eq.~(\ref{eq-diffusion1}) to a 1-dimensional diffusion equation.

\begin{figure}
\includegraphics[height=6.5cm] {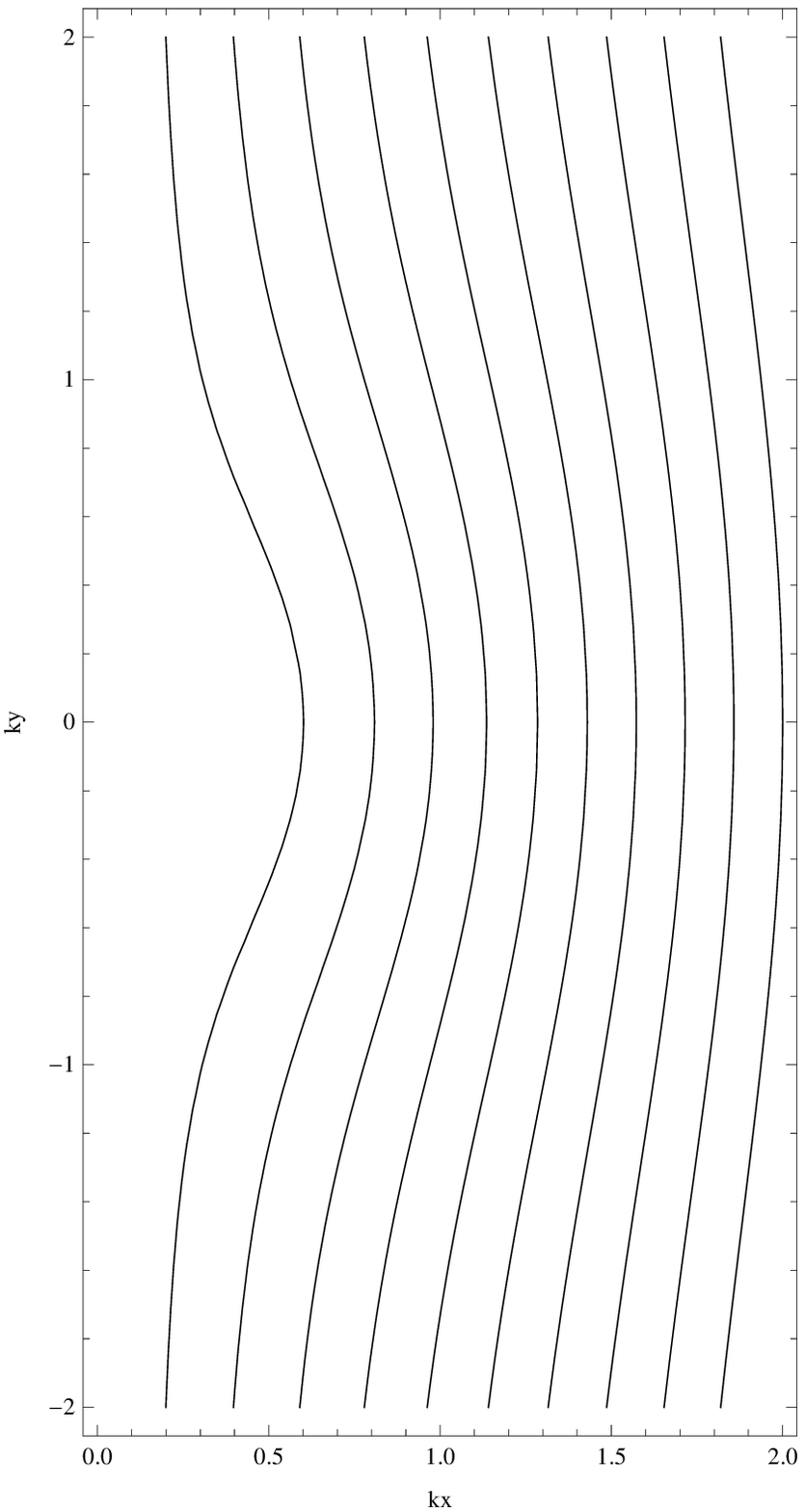}
\caption{\label{fig-OmegaCurves}}
The level sets of the function $\Omega_\k$ given by Eq.~(\ref{eq-W}).
\end{figure}

The general structure of Eq.~(\ref{eq-diffusion1}) is
\begin{equation}
\label{eq-diffusion2}
\pd{n_\k}{t} = \left(\pd{}{k_1},\pd{}{k_2}\right) S(\k) \left( \begin{array}{c}\pd{}{k_1}\\\pd{}{k_2}\end{array}\right) \ n_\k 
\end{equation}
Under a change of coordinates,
\begin{eqnarray}
\label{eq-varChange} k_1 &\to& \kappa_1(k_1,k_2)\\
\nonumber k_2 &\to& \kappa_2(k_1,k_2),
\end{eqnarray}
Eq.~(\ref{eq-diffusion2}) transforms to \cite{SMI1934} 
\begin{widetext}
\begin{equation}
\label{eq-changeOfCoordinates} \pd{n_\k}{t} = \left|\det J\right| \left(\pd{}{\kappa_1},\pd{}{\kappa_2}\right) \left|\det J\right|^{-1} J\, S(\k)\, J^T \left( \begin{array}{c}\pd{}{\kappa_1}\\\pd{}{\kappa_2}\end{array}\right) \ n_\k ,
\end{equation}
\end{widetext}
where our notation tacitly assumes that Eq.~(\ref{eq-varChange}) is used to express
all $\k$ dependence in  $S(\k)$ and $n_\k$ in terms of $\kappa_1$ and $\kappa_2$. 
$J$ is the Jacobian matrix of the change of variables:
\begin{equation}
J = \left(\begin{array}{cc}\pd{\kappa_1}{k_1}&\pd{\kappa_1}{k_2}\\\pd{\kappa_2}{k_1}&\pd{\kappa_2}{k_2}\end{array} \right).
\end{equation}
For a typical anisotropic diffusion equation for which the diffusion tensor would
have two positive eigenvalues, the maximum simplification would be obtained if new
coordinates could be found so that $J\, S(\k)\, J^T$ is diagonal. In this case, 
since the diffusion tensor is singular, we can do better and reduce the equation 
to a one-dimensional diffusion equation. By direct computation we observe that
\begin{eqnarray*}
&&\left(\begin{array}{cc}a&b\\c&d\end{array} \right)\, \left(\begin{array}{cc}\theta_\k^2&-\theta_\k\\-\theta_\k&1\end{array} \right) \, \left(\begin{array}{cc}a&b\\c&d\end{array} \right)^T\\
&& = \left(\begin{array}{cc}(b-a\, \theta_\k)^2&(b-a\,\theta_\k)(d-c\,\theta_\k)\\(b-a\,\theta_\k)(d-c\,\theta_\k)&(d-c\,\theta_\k)^2\end{array} \right).
\end{eqnarray*}
Using this we see that
\begin{displaymath}
J \left(\begin{array}{cc}\theta_\k^2&-\theta_\k\\-\theta_\k&1\end{array} \right) J^T = \left(\begin{array}{cc}1&0\\0&0\end{array} \right)
\end{displaymath}
provided that our new coordinates satisfy the equations
\begin{eqnarray*}
\label{eq-q1Eqn}\pd{\kappa_1}{k_2} - \theta_\k\,\pd{\kappa_1}{k_1} &=& 1\\
\label{eq-q2Eqn} \pd{\kappa_2}{k_2} - \theta_\k\, \pd{\kappa_2}{k_1} &=& 0.
\end{eqnarray*}
The first of these equations can be easily solved by inspection. The
second is also obvious once one recalls the definition of $\theta_\k$
from Eq.~(\ref{eq-theta}). We obtain:
\begin{eqnarray}
\label{eq-varChange2}\kappa_1(k_1,k_2) &=& k_2\\
\nonumber \kappa_2(k_1,k_2) &=& \W(k_1,k_2).
\end{eqnarray}
For this change of variables, $\det J = \pd{\W_\k}{k_1}$ so that in the 
$(\kappa_1,\kappa_2)$ plane, the diffusion equation Eq.~(\ref{eq-diffusion2}) takes the form
\begin{equation}
\label{eq-diffusion3}
\pd{n_\k}{t} = \pd{\W_\k}{k_1} \pd{}{\kappa_1} \tilde{S}(\k) \pd{n_\vv{\kappa}}{\kappa_1}.
\end{equation}
In the $(\kappa_1,\kappa_2)$ plane Eq.~(\ref{eq-diffusion3}) describes one dimensional 
diffusion in the $\kappa_1$ direction with $\kappa_2$ constant. Using 
Eqs.~(\ref{eq-varChange2}) to translate this back into the $(k_1,k_2)$ plane, 
Eq.~(\ref{eq-diffusion2}) therefore describes one-dimensional diffusion in the 
$k_2$ direction along lines of constant $\W_\k$. Illustrative examples
of the curves $\W_\k=$constant are plotted in Fig.~\ref{fig-OmegaCurves}.

\section{Nonlocal interaction with small scale zonal flows\label{sec-smallScaleZF}}

In this section we perform the analogous calculation assuming that the evolution
is dominated by interaction with the point $P_2=(0,2 k_2)$ corresponding to
small scale zonal flows. The results of this section were first
presented in \cite{NAZ1991}.

Following the same reasoning as before, we introduce a reference scale, $K$,
satisfying $K \ll k$ and $K\ll 1/\rho$, and neglect the contributions to
the collision integral coming from outside of a ball of radius $K$
of the point $\p=(0, 2\,k_2)$. We thus approximate Eq.~(\ref{eq-KE2}) by
\begin{eqnarray}
\nonumber \pd{n_\k}{t}&=& \int_{\left|\p-\q\right|<K} d\q\,F(\k,\q)\\
\label{eq-ssKE1} &=&-\int_{\left|\p-\q\right|<K} d\q\,F(\k-\q,-\q).
\end{eqnarray}
Taylor expanding the first argument about $\q=\p$ gives
\begin{widetext}
\begin{eqnarray}
\nonumber \pd{n_\k}{t}&=& -\int_{\left|\p-\q\right|<K} d\q\left[F(\k-\p,-\q) + (\q-\p)\cdot\nabla_{\q_1} \left. F(\k-\q_1,\q)\right|_{\q_1=\p} + O(\left|\p-\q\right|^2)\right]\\
\label{eq-ssKE2} &\approx& = -\int_{\left|\p-\q\right|<K} d\q\left[ F(\k-\p,-\q) - (q_1, q_2-2\,k_2) \cdot
\left(\partial_{k_1}F(\tilde{\k},-\q), \partial_{k_2}F(\tilde{\k},-\q)\right)\right]
\end{eqnarray}
where we have introduced the shorthand notation
\begin{equation}
\tilde{\k} = \k-\p = (k_1,-k_2).
\end{equation}
We now expand $F(\tilde{\k},-\q)$ about $\q=\p$. Since $V_{\q\,\k\!-\!\q\,\k}$
and $n_\q$ both vary rapidly near $\q=\p$, we should only expand the
$n_{\tilde{\k}+\q}-n_{\tilde{\k}}$ and $\delta(\W_{\tilde{\k}}-\W_{-\q}-\W_{\tilde{\k}+\q})$ terms. Let us look at these two terms in turn.
\begin{eqnarray}
\nonumber n_{\tilde{\k}+\q}-n_{\tilde{\k}} &=& n_{\tilde{\k}+\p}-n_{\tilde{\k}}
+ (\q-\p)\cdot \nabla_{\q^\prime}\left. n(\tilde{\k}+\q^\prime)\right|_{\q^\prime=\p} + O(\left|\p-\q\right|^2)\\
\label{eq-TaylorF}&\approx& n_{\k}-n_{\tilde{\k}} + (q_1, q_2-2\,k_2) \cdot \left(\partial_{k_1}n_\k,\partial_{k_2}n_\k\right).
\end{eqnarray}
Note that, unlike the expansion of the collision integral about the point
$(0,0)$ corresponding to large scale zonal flows detailed in 
Sec.~\ref{sec-largeScaleZF}, the leading order term in the Taylor expansion of
$F(\tilde{\k},-\q)$ about $\q=\p$ is not necessarily zero. It vanishes only if the
spectrum is symmetric about the $k_1$ axis (ie $n(k_1,k_2)=n(k_1,-k_2)$). We shall
return to this point below. Next, let us look at the argument of the 
$\delta$-function near $\q=\p$.
\begin{eqnarray}
\nonumber \W_{\tilde{\k}}-\W_{-\q}-\W_{\tilde{\k}+\q}  &=& \W_{\tilde{\k}}-\W_{-\p}-\W_{\k} + q_1\left[\pd{\W}{q^\prime_1}(-q^\prime_{1},-q^\prime_{2}) + \pd{\W}{q^\prime_{1}}(k_1+q^\prime_{1},-k_2+q^\prime_{2}) \right]_{(q^\prime_{1},q^\prime_{2}) = (0, 2\,k_2)}\\
\nonumber & & + (q_2-2\,k_2)\,\left[\pd{\W}{q^\prime_{2}}(-q^\prime_{1},-q^\prime_{2}) + \pd{\W}{q^\prime_{2}}(k_1+q^\prime_{1},-k_2+q^\prime_{2}) \right]_{(q^\prime_{1},q^\prime_{2}) = (0, 2\,k_2)} + O(\left|\p-\q\right|^2)\\
&\approx& -\pd{\W_\k}{k_2}\left( q_2 - 2\,k_2+\xi_\k\,q_1\right),
\end{eqnarray}
where
\begin{equation}
\label{eq-xi}
\xi_\k = \frac{3(k_1^2 -k_2^2) + \rho^2(k_1^4+6k_1^2k_2^2-3k_2^4)}{2 k_1 k_2 (1+4\rho^2k_2^2)}.
\end{equation}
The various derivatives have been computed from Eq.~(\ref{eq-W}). For example: 
\begin{displaymath}
\left[ \pd{\W}{q^\prime_{2}}(k_1+q^\prime_{1},-k_2+q^\prime_{2})\right]_{(q^\prime_{1},q^\prime_{2}) = (0, 2\,k_2)} = \pd{\W_\k}{k_2} =-\frac{2 \beta \rho^4 k_1k_2}{(1+\rho^2 k^2)^2},
\end{displaymath}
and so forth.  The leading order term is zero owing to the fact that $\k,\p$ and 
$\tilde{\k}$ are resonant. Putting this together, the Taylor expansion of the 
$\delta$-function is
\begin{equation}
\label{eq-TaylorDeltaFn}
\delta(\W_{\tilde{\k}}-\W_{-\q}-\W_{\tilde{\k}+\q}) \approx \left| \pd{\W_\k}{k_2} \right|^{-1}\,\delta \left( q_2 - 2\,k_2+\xi_\k\,q_1\right).
\end{equation}
From Eqs.~(\ref{eq-ssKE2}), (\ref{eq-TaylorF}) and (\ref{eq-TaylorDeltaFn}) we
see that the leading order term in the kinetic equation coming from nonlocal
interaction with small scale zonal flows is:
\begin{equation}
\label{eq-ssKE3}
\pd{n_\k}{t} = Y_\k\, \left[n(k_1,-k_2)-n(k_1,k_2)\right],
\end{equation}
where
\begin{eqnarray}
\nonumber Y_\k &=& 4\pi \left| \pd{\W_\k}{k_2} \right|^{-1} \int_{\left|\p-\q\right|<K} dq_1 dq_2  \left|V_{-\q\,\k\!+\!\q\,\tilde{\k}}\right|^2 \,\delta(q_2 - 2\,k_2+\xi_\k\,q_1)\, n(q_1,q_2)\\
\nonumber &=& 4\pi \left| \pd{\W_\k}{k_2} \right|^{-1} \int_{\left|\p-\q\right|<K} dq_1 dq_2  \left|-V_{\q\,\k\!-\!\q\,\k}\right|^2 \,\delta(q_2 - 2\,k_2-\xi_\k\,q_1)\, n(-q_1,q_2)\\
\label{eq-Yk} &=& 4\pi \left| \pd{\W_\k}{k_2} \right|^{-1} \int_{\left|q_1\right|<K} \left[\left|V_{\q\,\k\!-\!\q\,\k}\right|^2 n_\q\right]_{q_2=2\,k_2} dq_1.
\end{eqnarray}
In the intermediate step, we have used the symmetries, Eq.~(\ref{eq-Vsymmetries}), 
of $V^\q_{\k\!-\!\q\,\k}$ and relabelled the integration variable $q_1 \to -q_1$
to bring the interaction coefficient to a neater form. At the final step the
delta function has been used to integrate out $q_2$ to leading order. We have
also assumed that the spectrum is symmetric about the $k_2$ axis although this
is an inessential point. Eq.~(\ref{eq-ssKE3}) tells us that the leading order 
effect of interactions with small scale zonal flows is to cause the spectrum
to relax to a state which is symmetric about the $k_1$ axis, a point which
was first made in \cite{BNZ1990,BNZ1990B}.

If we wish to observe any redistribution of spectral energy density due to
the interactions with small scale zonal flows, we need to consider the higher
order terms in Eq.~(\ref{eq-ssKE2}). The first order term vanishes after 
integration over $\q$ since the integrand is an odd function of $\p-\q$. The
next contribution is therefore the second order one. From Eqs.~(\ref{eq-ssKE2})
and (\ref{eq-TaylorF}) the second order contribution can again be
presented as an anisotropic diffusion equation in $\k$-space:
\begin{equation}
\label{eq-ssDiffusion1}
\pd{n_\k}{t} = \pd{}{k_i}\,B_{ij}(\k) \pd{n_\k}{k_j}
\end{equation}
where the diffusion tensor is given by the matrix
\begin{eqnarray}
\nonumber B(\k) &=& 4 \pi \left| \pd{\W_\k}{k_2} \right|^{-1} \int_{\left|\p-\q\right|<K} dq_1 dq_2  \left|V_{-\q\,\k\!+\!\q\,\tilde{\k}}\right|^2 \,\delta(q_2 - 2\,k_2+\xi_\k\,q_1)\, n(q_1,q_2) \left(\begin{array}{cc}q_1^2&q_1(q_2-2\,k_2)\\q_1(q_2-2\,k_2)&(q_2-2\,k_2)^2\end{array} \right)\\
\nonumber&=& 4 \pi \left| \pd{\W_\k}{k_2} \right|^{-1} \int_{\left|\p-\q\right|<K} dq_1 dq_2  \left|-V_{\q\,\k\!-\!\q\,\k}\right|^2 \,\delta(q_2 - 2\,k_2-\xi_\k\,q_1)\, n(-q_1,q_2) \left(\begin{array}{cc}q_1^2&-q_1(q_2-2\,k_2)\\-q_1(q_2-2\,k_2)&(q_2-2\,k_2)^2\end{array} \right)\\
\nonumber &=& 4 \pi \left| \pd{\W_\k}{k_2} \right|^{-1} \left(\begin{array}{cc}1&-\xi_\k\\-\xi_\k&\xi_\k^2\end{array} \right) \int_{\left|q_1\right|<K} \left[ \left|V_{\q\,\k\!-\!\q\,\k}\right|^2 n(-q_1,q_2)\, q_1^2\right]_{q_2=2\,k_2} dq_1 \\
\label{eq-B}&=& \left| \pd{\W_\k}{k_2} \right|^{-1}\, \tilde{B}(\k)\, \left(\begin{array}{cc}1&-\xi_\k\\-\xi_\k&\xi_\k^2\end{array} \right)  
\end{eqnarray}
where we have performed the same manipulations as those used to arrive at
Eq.~(\ref{eq-Yk}) above and defined the scalar quantity
\begin{equation}
\label{eq-Btilde}
\tilde{B}(\k) = 4\pi\,\int_{\left|q_1\right|<K} \left[ \left|V_{\q\,\k\!-\!\q\,\k}\right|^2 n(-q_1,q_2)\, q_1^2\right]_{q_2=2\,k_2} dq_1.
\end{equation}
The diffusion tensor, $B(\k)$, is again singular
indicating that Eq.~(\ref{eq-ssDiffusion1}) should be reducible to
a one-dimensional diffusion equation by an appropriate change of
variables. Following the same proceedure as in Sec.~\ref{sec-largeScaleZF}, we
introduce new variables
\begin{eqnarray}
\label{eq-ssVarChange2} k_1 &\to& \kappa_1(k_1,k_2)\\
\nonumber k_2 &\to& \kappa_2(k_1,k_2),
\end{eqnarray}
whose Jacobian, $J$, given by
\begin{displaymath}
J = \left(\begin{array}{cc}\pd{\kappa_1}{k_1}&\pd{\kappa_1}{k_2}\\\pd{\kappa_2}{k_1}&\pd{\kappa_2}{k_2}\end{array} \right),
\end{displaymath}
should diagonalise the matrix 
$\left(\begin{array}{cc}1&-\xi_\k\\-\xi_\k&\xi_\k^2\end{array} \right)$ appearing
in the diffusion tensor, Eq.~(\ref{eq-B}). Observing that
\begin{displaymath}
\left(\begin{array}{cc}a&b\\c&d\end{array} \right)\, \left(\begin{array}{cc}1&-\xi_\k\\-\xi_\k&\xi_\k^2\end{array} \right) \, \left(\begin{array}{cc}a&b\\c&d\end{array} \right)^T = \left(\begin{array}{cc}(a-b\, \xi_\k)^2&(a-b\,\xi_\k)(c-d\,\xi_\k)\\(a-b\,\xi_\k)(c-d\,\xi_\k)&(c-d\,\xi_\k)^2\end{array} \right),
\end{displaymath}
we see that
\begin{displaymath}
J \left(\begin{array}{cc}1&-\xi_\k\\-\xi_\k&\xi_\k^2\end{array} \right) J^T = \left(\begin{array}{cc}1&0\\0&0\end{array} \right)
\end{displaymath}
provided that our new coordinates satisfy the equations
\begin{eqnarray*}
\label{eq-kappa1Eqn}\pd{\kappa_1}{k_1} - \xi_\k\,\pd{\kappa_1}{k_2} &=& 1\\
\label{eq-kappa2Eqn} \pd{\kappa_2}{k_1} - \xi_\k\, \pd{\kappa_2}{k_2} &=& 0.
\end{eqnarray*}
The first equation can be easily solved by inspection while the second requires a 
little more effort using the method of characteristics. One obtains
\begin{eqnarray}
\label{eq-kappa1}\kappa_1(\k) &=& k_1\\
\nonumber \kappa_2(\k) = Z_\k &=& \arctan \left( \frac{k_2 + \sqrt{3}\, k_1}{\rho\, k^2}\right)
-\arctan \left( \frac{k_2 - \sqrt{3}\, k_1}{\rho\, k^2}\right)\\
\label{eq-Z} &&-\frac{2\sqrt{3}\,\rho\,k_1}{1+\rho^2\,k^2}.
\end{eqnarray}
With this done, $\det J = \pd{Z_\k}{k_2}$, so that, 
according to Eq.~(\ref{eq-changeOfCoordinates}), the diffusion equation, 
Eq.~(\ref{eq-ssDiffusion1}), transforms into
\begin{equation}
\label{eq-ssDiffusion2}
\pd{n_\vv{\kappa}}{t} =   \pd{Z_\k}{k_2} \pd{}{\kappa_1} \left[ \left( \pd{Z_\k}{k_2} \right)^{-1} \left( \pd{\W_\k}{k_2} \right)^{-1}\, \tilde{B}(\k)\, \pd{n_\vv{\kappa}}{\kappa_1}\right].
\end{equation}
where $\tilde{B}(\k)$ is given by Eq.~(\ref{eq-Btilde}).
In the $(\kappa_1,\kappa_2)$ plane Eq.~(\ref{eq-ssDiffusion2}) describes one 
dimensional diffusion in the $\kappa_1$ direction with $\kappa_2$ constant. 
Translating back into the $(k_1,k_2)$ plane, using Eqs.~(\ref{eq-kappa1})
and (\ref{eq-Z}), Eq.~(\ref{eq-ssDiffusion2}) therefore describes one-dimensional 
diffusion in the $k_1$ direction along lines of constant $Z_\k$. Illustrative 
examples of the curves $Z_\k=$constant are plotted in Fig.~\ref{fig-musselCurves}.
\end{widetext}

\begin{figure}
\includegraphics[height=6.5cm] {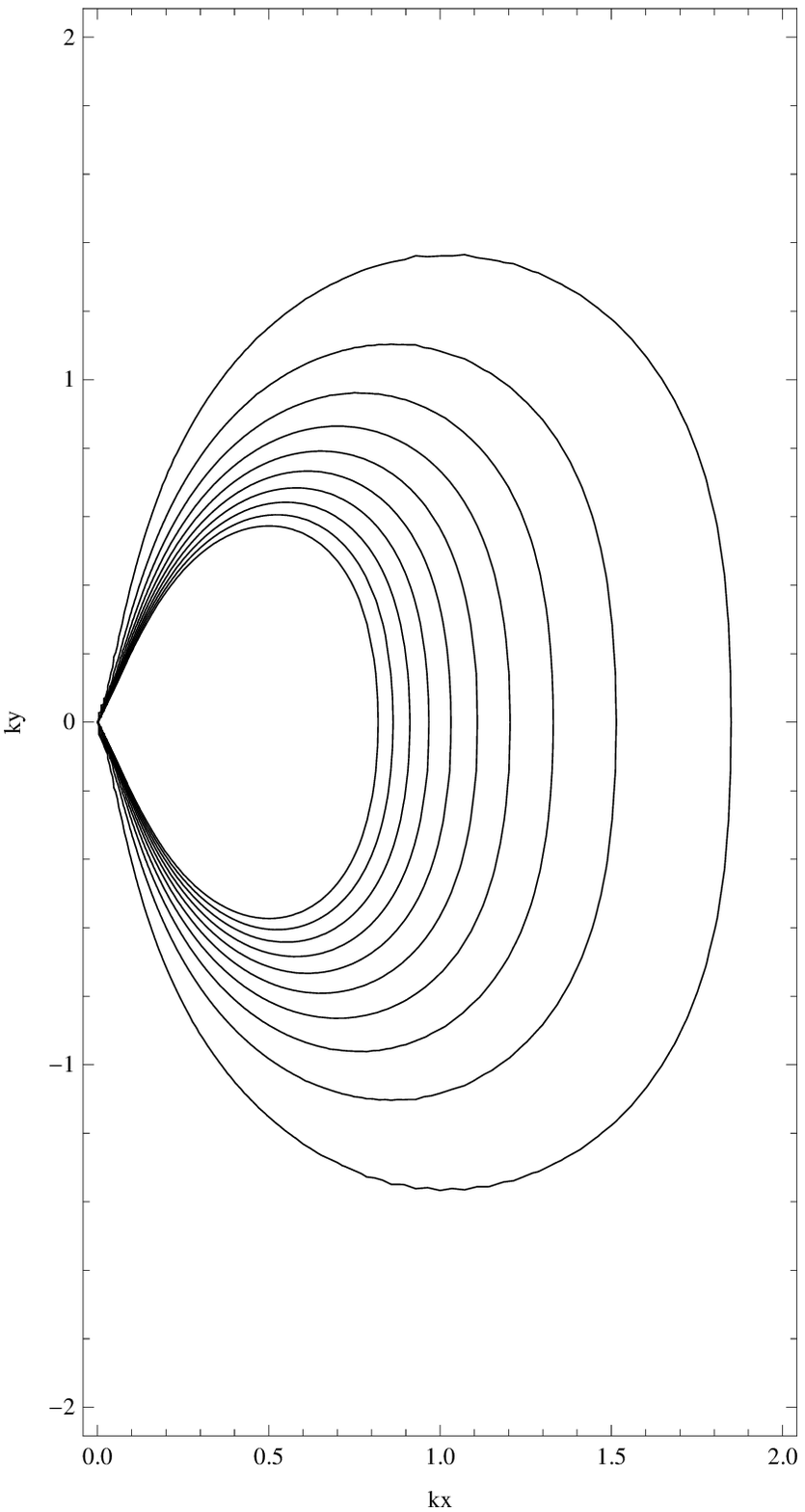}
\caption{\label{fig-musselCurves}}
The level sets of the function $Z_\k $ given by Eq.~(\ref{eq-Z}).
\end{figure}

Interestingly $Z_\k$ defined in Eq.~(\ref{eq-Z}), is closely related to the third 
invariant (in additional to the energy and enstrophy) of the full kinetic 
equation, Eq.~(\ref{eq-KE}), discovered in \cite{BNZ1991} 
(see also \cite{BAL1991,BAL1997}). From Fig. \ref{fig-musselCurves} we note that
the diffusion due to interaction with small scale zonal flows tends to
transfer the wave action to large scales, a process which was investigated
numerically in \cite{NQ2009}.

\section{Conclusions}

In this note we have summarised the derivation of the nonlocal kinetic
equation for Rossby wave turbulence modelled by the wave kinetic equation
obtained from the Charney-Hasegawa-Mima equation. For small scale turbulence
there are two possible sources of nonlocal interactions in
$\k$-space. The first is with large scale zonal flows with spectral support around 
the point $P_1=(0,0)$. The second is with small scale zonal flows with spectral 
support around the point $P_2=(0,2\,k_2)$. In both cases, the redistribution of
small-scale wave action in $\k$-space is described at leading order by an
anisotropic diffusion equation in $\k$-space. In both cases, the diffusion
tensor is singular meaning that, via an appropriate change of variables, it
can be shown that the diffusion occurs along one-dimensional curves in
$\k$-space. We showed how to obtain these curves explicitly. In the case of nonlocal
interaction with the point $P_1$, the curves are open and tend to
transport waveaction to $k_2=\pm\infty$. In the case of nonlocal interaction
with the point $P_2$, the curves are closed and tend to transport wave action
towards the origin. Recent numerical investigations have demonstrated the
presence of these diffusive mechanisms in simulations of the original 
CHM equation \cite{CNQ2010}.

\bibliography{diffusion}

\begin{thebibliography}{11}
\expandafter\ifx\csname natexlab\endcsname\relax\def\natexlab#1{#1}\fi
\expandafter\ifx\csname bibnamefont\endcsname\relax
  \def\bibnamefont#1{#1}\fi
\expandafter\ifx\csname bibfnamefont\endcsname\relax
  \def\bibfnamefont#1{#1}\fi
\expandafter\ifx\csname citenamefont\endcsname\relax
  \def\citenamefont#1{#1}\fi
\expandafter\ifx\csname url\endcsname\relax
  \def\url#1{\texttt{#1}}\fi
\expandafter\ifx\csname urlprefix\endcsname\relax\def\urlprefix{URL }\fi
\providecommand{\bibinfo}[2]{#2}
\providecommand{\eprint}[2][]{\url{#2}}

\bibitem[{\citenamefont{Zakharov et~al.}(1992)\citenamefont{Zakharov, Lvov, and
  Falkovich}}]{ZLF92}
\bibinfo{author}{\bibfnamefont{V.}~\bibnamefont{Zakharov}},
  \bibinfo{author}{\bibfnamefont{V.}~\bibnamefont{Lvov}}, \bibnamefont{and}
  \bibinfo{author}{\bibfnamefont{G.}~\bibnamefont{Falkovich}},
  \emph{\bibinfo{title}{Kolmogorov Spectra of Turbulence}}
  (\bibinfo{publisher}{Springer-Verlag}, \bibinfo{address}{Berlin},
  \bibinfo{year}{1992}).

\bibitem[{\citenamefont{Newell et~al.}(2001)\citenamefont{Newell, Nazarenko,
  and Biven}}]{NNB01}
\bibinfo{author}{\bibfnamefont{A.}~\bibnamefont{Newell}},
  \bibinfo{author}{\bibfnamefont{S.}~\bibnamefont{Nazarenko}},
  \bibnamefont{and} \bibinfo{author}{\bibfnamefont{L.}~\bibnamefont{Biven}},
  \bibinfo{journal}{Physica D} \textbf{\bibinfo{volume}{152-153}},
  \bibinfo{pages}{520} (\bibinfo{year}{2001}).

\bibitem[{\citenamefont{{Balk} et~al.}(1990{\natexlab{a}})\citenamefont{{Balk},
  {Nazarenko}, and {Zakharov}}}]{BNZ1990}
\bibinfo{author}{\bibfnamefont{A.~M.} \bibnamefont{{Balk}}},
  \bibinfo{author}{\bibfnamefont{S.~V.} \bibnamefont{{Nazarenko}}},
  \bibnamefont{and} \bibinfo{author}{\bibfnamefont{V.~E.}
  \bibnamefont{{Zakharov}}}, \bibinfo{journal}{Phys. Lett. A}
  \textbf{\bibinfo{volume}{146}}, \bibinfo{pages}{217}
  (\bibinfo{year}{1990}{\natexlab{a}}).

\bibitem[{\citenamefont{{Balk} et~al.}(1990{\natexlab{b}})\citenamefont{{Balk},
  {Nazarenko}, and {Zakharov}}}]{BNZ1990B}
\bibinfo{author}{\bibfnamefont{A.~M.} \bibnamefont{{Balk}}},
  \bibinfo{author}{\bibfnamefont{S.~V.} \bibnamefont{{Nazarenko}}},
  \bibnamefont{and} \bibinfo{author}{\bibfnamefont{V.~E.}
  \bibnamefont{{Zakharov}}}, \bibinfo{journal}{Sov. Phys. - JETP}
  \textbf{\bibinfo{volume}{71}}, \bibinfo{pages}{249}
  (\bibinfo{year}{1990}{\natexlab{b}}).

\bibitem[{\citenamefont{Smith}(1934)}]{SMI1934}
\bibinfo{author}{\bibfnamefont{T.}~\bibnamefont{Smith}},
  \bibinfo{journal}{Proc. Phys. Soc.} \textbf{\bibinfo{volume}{34}},
  \bibinfo{pages}{344 } (\bibinfo{year}{1934}).

\bibitem[{\citenamefont{{Nazarenko, S.V.}}(1991)}]{NAZ1991}
\bibinfo{author}{\bibnamefont{{Nazarenko, S.V.}}}, \bibinfo{journal}{JETP
  Lett.} \textbf{\bibinfo{volume}{53}}, \bibinfo{pages}{628}
  (\bibinfo{year}{1991}).

\bibitem[{\citenamefont{{Balk} et~al.}(1991)\citenamefont{{Balk}, {Nazarenko},
  and {Zakharov}}}]{BNZ1991}
\bibinfo{author}{\bibfnamefont{A.~M.} \bibnamefont{{Balk}}},
  \bibinfo{author}{\bibfnamefont{S.~V.} \bibnamefont{{Nazarenko}}},
  \bibnamefont{and} \bibinfo{author}{\bibfnamefont{V.~E.}
  \bibnamefont{{Zakharov}}}, \bibinfo{journal}{Phys. Lett. A}
  \textbf{\bibinfo{volume}{152}}, \bibinfo{pages}{276} (\bibinfo{year}{1991}).

\bibitem[{\citenamefont{{Balk}}(1991)}]{BAL1991}
\bibinfo{author}{\bibfnamefont{A.~M.} \bibnamefont{{Balk}}},
  \bibinfo{journal}{Phys. Lett. A} \textbf{\bibinfo{volume}{155}},
  \bibinfo{pages}{20} (\bibinfo{year}{1991}).

\bibitem[{\citenamefont{{Balk}}(1997)}]{BAL1997}
\bibinfo{author}{\bibfnamefont{A.~M.} \bibnamefont{{Balk}}},
  \bibinfo{journal}{SIAM Review} \textbf{\bibinfo{volume}{39}},
  \bibinfo{pages}{68} (\bibinfo{year}{1997}).

\bibitem[{\citenamefont{Nazarenko and Quinn}(2009)}]{NQ2009}
\bibinfo{author}{\bibfnamefont{S.}~\bibnamefont{Nazarenko}} \bibnamefont{and}
  \bibinfo{author}{\bibfnamefont{B.}~\bibnamefont{Quinn}},
  \bibinfo{journal}{Phys. Rev. Lett.} \textbf{\bibinfo{volume}{103}},
  \bibinfo{pages}{118501} (\bibinfo{year}{2009}).

\bibitem[{\citenamefont{{Connaughton} et~al.}(2010)\citenamefont{{Connaughton},
  {Nazarenko}, and {Quinn}}}]{CNQ2010}
\bibinfo{author}{\bibfnamefont{C.}~\bibnamefont{{Connaughton}}},
  \bibinfo{author}{\bibfnamefont{S.~V.} \bibnamefont{{Nazarenko}}},
  \bibnamefont{and} \bibinfo{author}{\bibfnamefont{B.~E.}
  \bibnamefont{{Quinn}}} (\bibinfo{year}{2010}),
  \bibinfo{note}{arXiv:1008.3338v1 [nlin.CD]}.

\end{thebibliography}
\end{document}